\documentclass[conference,usenatbib]{basi}
\usepackage[T1]{fontenc}
\usepackage[british]{babel}
\usepackage[varg]{txfonts}
\usepackage{rotating}
\usepackage{dcolumn}
\begin{document}
\title[``Blobs'' in blazar jets]{``Blobs'' in blazar jets}
\author[Prasad Subramanian]%
       {Prasad Subramanian\thanks{email: \texttt{p.subramanian@iiserpune.ac.in}}
       \\ Indian Institute of Science Education and Research, Pune}

\pubyear{2015}
\volume{12}
\pagerange{\pageref{firstpage}--\pageref{lastpage}}

\date{Received --- ; accepted ---}

\maketitle
\label{firstpage}

\begin{abstract}
The concept of highly relativistic electrons confined to blobs that are moving out with modestly relativistic speeds is often invoked to explain high energy blazar observations. The important parameters in this model such as the bulk Lorentz factor of the blob ($\Gamma$), the random Lorentz factor of the electrons ($\gamma$) and the blob size are typically observationally constrained, but its not clear how and why the energetic electrons are held together as a blob. Here we present some preliminary ideas based on scenarios for cosmic ray electron self-confinement that could lead to a coherent picture.
%
%
\end{abstract}

\begin{keywords}
   Galaxies: active--jets
\end{keywords}

\section{Introduction}\label{s:intro}
The general narrative in blazar physics concentrates on energetic electrons with random Lorentz factors ($\gamma$) ranging from around $10^4$ to $10^6$ radiating via either the synchtroton process and/or the inverse compton mechanism, in order to explain observations. In order to account for the high apparent luminosities, its generally hypothesized that the energetic electrons are also moving outwards with a bulk Lorentz factor ($\Gamma$) ranging from 10 to 50. In particular, energetic electrons are thought to be confined to {\em blobs}, which move outwards with the jet bulk flow. The quantities $\gamma$ and $\Gamma$ and the blob size are generally constrained using observations, but its not clear how the energetic electrons are confined to the blob - and for that matter, what the blob really is.
\section{Confinement of electrons in the blob}
The transverse dimension of the blob is determined by the variability timescale (using the light crossing argument), which we take to be 1 day for the sake of concreteness. It is well known from cosmic ray research that energetic electrons cannot stream down a large-scale magnetic field; the anisotropy of the electron beam leads to a resonant Alfv\'en wave instability causes them to shed a turbulent spectrum of hydromagnetic waves (Kulsrud \& Pearce 1969; Wentzel 1969; Ceasrsky \& Kuslrud 1973). Pitch angle scattering off this turbulent wave spectrum isotropizes the electron beam and causes them to be confined or ``bunched up'' into a blob - this is a mechanism of self-confinement. The crucial question is - what is the effective mean free path for this self-confinement mechanism, and how does it compare with a typical blob size? Early research on cosmic rays only required cosmic rays to be confined to the galactic disk, and the predicted mean free paths were therefore fairly large. However, recent hybrid particle-MHD simulations have suggested that a non-resonant electron beam-driven instability can result in a self-confinement mean free path $\approx$ an electron Larmor radius (Reville et al 2008). For magnetic fields ranging from 1 to 100 Gauss (which are typical values invoked to fit observational data), and for $\gamma = 10^4$, the electron Larmor radius is somewhat smaller than (around 0.1 to 0.4) a blob size corresponding to a variability timescale of 1 day. In general, the larger the magnetic field, the larger the $\Gamma$, the longer the variability timescale and the smaller the $\gamma$, more Larmor radii ``fit inside'' a blob; in other words, a blob could be an aggregation of several self-confinement mean free paths. 

\section{Conclusion}
Energetic electrons accelerated to form a jet-like beam are prone to shedding a spectrum of hydromagnetic waves, which prevents them from streaming at speeds greater than the local Alfv\'en speed. The electrons are likely to undergo pitch angle scattering (and consequent isotropization) off these hydromagnetic waves, leading them to be spatially self-confined/bunched up into blobs. The timescale for pitch angle scattering is typically $\ll$ relevant energy loss timescales. Although further work is required, initial results seem to be quite encouraging. As for cosmic rays, the hydromagnetic wave spectrum could also act as the ``clutch'' that couples the energetic electrons in the blob to the bulk jet flow (Wentzel 1971).











\end{document}